\documentclass[aps,prd,preprintnumbers,twocolumn]{revtex4}
\usepackage{amsmath}
\usepackage{graphicx}
\usepackage{dsfont}
\usepackage{subfigure}
\usepackage{epsfig}
\usepackage{epsf}

\def\nab{\nabla}
\def\3nab{\tilde{\nabla}}

\def\nn{\nonumber}

\def\hsp5{\hspace{5mm}}

\def\case#1/#2{\textstyle\frac{#1}{#2}}

\def\be {\begin{equation}}
\def\ee {\end{equation}}
\def\bea {\begin{eqnarray}}
\def\eea {\end{eqnarray}}
\def\ber{\begin{eqnarray*}}
\def\ear{\end{eqnarray*}}

\def\bc {\begin{center}}
\def\ec {\end{center}}
\def\case#1/#2{\frac{#1}{#2}}
\def\rf#1{(\ref{#1})}
 \newcommand{\na}{\nabla}

\newcommand{\bra}[1]{\left(#1\right)}
\newcommand{\bras}[1]{\left[#1\right]}

\newcommand \veps {\varepsilon} %curly epsilon
\newcommand{\A}{{\cal A}}
\newcommand{\E}{{\cal E}}
\renewcommand{\H}{{\cal H}}
\newcommand{\LB}{\left(}
\newcommand{\RB}{\right)}
\newcommand{\lc}{\varepsilon}

\def\etal\;{{\it et al.}}

\begin{document}
%%%%%%%%%%%%%%%%%%%%%%%%%%%%%%%%%%%%%%%%%%%%%%%%%%%%
\title{A Geometrical Approach to Strong Gravitational Lensing in f(R) Gravity}
%%%%%%%%%%%%%%%%%%%%%%%%%%%%%%%%%%%%%%%%%%%%%%%%%%%%
\author{Anne Marie Nzioki  \$ \dag,  Peter K.S. Dunsby  \$ \dag \ddag, Rituparno Goswami  \$ \dag $\;$
and Sante Carloni *}
\affiliation{\$  Centre for Astrophysics, Cosmology and Gravitation,  University of Cape Town, Rondebosch,
7701, South Africa}
\address{\dag \ Department of Mathematics and Applied Mathematics, 
University of Cape Town, Rondebosch,
7701, South Africa}
\address{\ddag \ South African Astronomical Observatory, 
Observatory, Cape Town, South Africa}
\address{* Institut d'Estudis Espacials de Catalunya
(IEEC), Campus UAB, Facultat Ci\`{e}ncies, Torre C5-Par-2a pl, E-08193 Bellaterra
(Barcelona) Spain,}
\date{\today}
%\eads{anne.nzioki@gmail.com, peter.dunsby@uct.ac.za, Rituparno.Goswami@uct.ac.za and %sante.carloni@gmail.com }
%%%%%%%%%%%%%%%%%%%%%%%%%%%%%%%%%%%%%%%%%%%%%%%%%%%%%

%%%%%%%%%%%%%%%%%%%%%%%%%%%%%%%%%%%%%%%%%%%%%%%%%%%%%
\begin{abstract}
We present a framework for the study of lensing in spherically symmetric spacetimes within the context of $f(R)$ gravity. 
Equations for the propagation of null geodesics, together with an expression for the bending angle are derived for any
$f(R)$ theory and then applied to an exact spherically symmetric solution of $R^n$ gravity. We find that for this case more bending is expected for  $R^{n}$ gravity theories in comparison to GR and is dependent on the value of $ n $ and the value of distance of closest approach of the incident null geodesic. 
\end{abstract}
 \maketitle
 
%%%%%%%%%%%%%%%%%%%%%%%%%%%%%%%%%%%%%%%%%
\section{Introduction}
%%%%%%%%%%%%%%%%%%%%%%%%%%%%%%%%%%%%%%%%% 
Modifications of the theory of General Relativity (GR) by including higher order corrections to the theory were first introduced by Herman Weyl \cite{bi:Weyl} as early as 1918,  just three years after Einstein completed his theory of General Relativity. Weyl proposed modifications to the theory by including higher order invariants in the action, in order to extend general relativity's geometrical foundations. In 1921, Arthur Eddington also began to consider fourth order theories of gravity and went on to publish his famous book: \textit{The Mathematical Theory of Relativity} \cite{bi:Eddington}, which contained work on generalized versions of Weyl's theory. Ever since, there has been a great number of gravitational theories proposing modifications to the Einstein-Hilbert action.

Originally the motivation for considering modified theories of gravity stems from limitations of General Relativity when considering strong gravity regimes. More recently however, corrections to GR have been proposed to account for the dark sector of the universe. Studies of the Cosmic Microwave Background (CMB) \cite{CMB}, Supernovae Type Ia surveys \cite{SN} and Baryon Acoustic Oscillations \cite{BAO} indicate that the energy density budget of the universe is $ 5 \% $ ordinary matter (baryons, radiation and neutrinos), $ 25\% $ dark matter and $ 70 \% $ dark energy. Dark matter is responsible for the gravitational clumping of galaxies, galaxy clusters and large-scale structures, whilst dark energy is linked to the current phase of accelerated expansion of the Universe. The $\Lambda$CDM model (\textit{concordance model}) is currently the best fit model, using the cosmological constant $ \Lambda $ to explain dark energy. An alternative explanation is to consider that on cosmological scales, the current description of gravity is incomplete and a modification to the theory is needed. One of the most popular modifications to gravity which has been proposed to explain the late-time acceleration of the Universe, without the need for dark energy, is $f(R)$ gravity, which represents a generalization of Einstein-Hilbert action by replacing the Ricci scalar $R$ with the function $f(R)$. These fourth order theories have been shown to and give rise to a phase of accelerated expansion without the need for dark energy \cite{bi:Capozziello2002,bi:Carroll, bi:Starobinsky, bi:Capozziello:2003tk} and to account for the rotation curves for spiral galaxies without the need for dark matter \cite{bi:Capozziello2006a}.

A good theory of gravity is clearly one that agrees with observations. One of the first experimental verifications of GR was the measurement of the bending of light, observed for the first time by Eddington in 1919 during a total solar eclipse. Since then, Gravitational lensing has been a powerful tool used to determine the mass distribution of galaxies and galaxy clusters and to put constraints on scales as small as stars to large scale structures and cosmological parameters \cite{bi:Ehlers}. Given that the lensing effect is dependent on the underlying theory of gravity, investigating modifications of GR would result in deviations from the standard expression of the deflection angle and is consequently worth investigating. The derivation of the form of the lensing angle for $f(R)$ theory has already been presented in \cite{bi:Capozziello2006b} and \cite{bi:Mendoza} using different methods.

In this paper we use covariant method to find general expressions for the propagation of null geodesics in static spherically symmetric spacetimes in $f(R)$ gravity theories. The solutions of these equations are then used to obtain a general expression for the deflection angle for spherically symmetric spacetimes. The outline of this paper is as follows. In section \ref{fourthorder} we present the general equations for $f(R)$ gravity. In section \ref{covariantapproach} we apply the1+1+2 covariant approach to static spherical symmetric spacetimes in $f(R)$ gravity. The propagation of null geodesics is considered in section \ref{lensgeometry} and then calculated explicitly for the case of $f(R)= R^{n}$ in static spherically symmetric spacetime in order to determine the bending angle in section \ref{lensapplication}.

Unless otherwise specified, natural units ($\hbar=c=k_{B}=8\pi G=1$) will be used throughout this paper, Latin indices run from 0 to 3. The symbol $\nabla$ represents the usual covariant derivative and $\partial$ corresponds to partial differentiation. We use the $-,+,+,+$ signature and the Riemann tensor is defined by
\begin{equation}
R^{a}{}_{bcd}=\Gamma^a{}_{bd,c}-\Gamma^a{}_{bc,d}+ \Gamma^e{}_{bd}\Gamma^a{}_{ce}-\Gamma^e{}_{bc}\Gamma^a{}_{de}\;,
\end{equation}
where the $\Gamma^a{}_{bd}$ are the Christoffel symbols (i.e. symmetric in the lower indices), defined by
\begin{equation}
\Gamma^a_{}{bd}=\frac{1}{2}g^{ae}
\left(g_{be,d}+g_{ed,b}-g_{bd,e}\right)\;.
\end{equation}
The Ricci tensor is obtained by contracting the {\em first} and the {\em third} indices
\begin{equation}\label{Ricci}
R_{ab}=g^{cd}R_{acbd}\;.
\end{equation}
The symmetrization and the antisymmetrization over the indexes of a tensor are defined as 
\begin{equation}
T_{(a b)}= \frac{1}{2}\left(T_{a b}+T_{b a}\right)\;,\qquad T_{[a b]}= \frac{1}{2}\left(T_{a b}-T_{b a}\right)\,.
\end{equation}
Finally the Hilbert--Einstein action in the presence of matter is given by
\begin{equation}
{\cal A}=\frac12\int d^4x \sqrt{-g}\left[R+ 2{\cal L}_m \right]\;.
\end{equation}

%%%%%%%%%%%%%%%%%%%%%%%%%%%%%%%%%%%%%%%%%
\section{General equations for fourth order gravity} \label{fourthorder}
%%%%%%%%%%%%%%%%%%%%%%%%%%%%%%%%%%%%%%%%% 
In a completely general context, a fourth order theory of gravity is obtained by adding the terms involving $f(R)$, $R_{ab}R^{ab}$, $R_{abcd}R^{abcd}$ to the standard Einstein Hilbert action. However, we know the Gauss-Bonnet term $(\mathcal{G}=R^2-4R_{ab}R^{ab}+ R_{abcd}R^{abcd})$ is a total differential in four dimensions and hence do not affect the field equations. Using this symmetry we can rewrite $R_{abcd}R^{abcd}$ in terms of the other two. Furthermore, if the spacetime is highly symmetric, then the variation of the term $R_{ab}R^{ab}$ can always be rewritten in terms of the variation of $R^2$ \cite{DeWitt:1965jb,Barth:1983hb}. It follows that the sufficiently general fourth-order Lagrangian for a highly symmetric spacetime only contains powers of $R$ and we can, with out loss of generality, write the action as 
\be
{\cal A}= \frac12 \int d^4x\sqrt{-g}\left[f(R) +{\cal L}_m\right]\;,
\label{action}
\ee
where ${\cal L}_m$ represents the matter contribution and $\Lambda$ is the usual cosmological constant.
Varying the action with respect to the metric gives the following field equations:
\be
f,_{R} G_{ab} = T^{m}_{ab}+ \frac12 (f-Rf,_{R}) 
+ \nabla_{b}\nabla_{a}f,_{R}- g_{ab}\nabla_{c}\nabla^{c}f,_{R}\,,
\label{field1} 
\ee
where $f,_{R}$ denotes the derivative of the function `$f$' w.r.t the Ricci scalar and $T^{m}_{ab}$ is the matter stress energy tensor defined as
\be
T^{m}_{ab} = \mu^{m}u_{a}u_{b} + p^{m}h_{ab}+ q^{m}_{a}u_{b}+ q^{m}_{b}u_{a}+\pi^{m}_{ab}\;.
\ee
Here $u^a$ is the direction of a timelike observer, $h_{ab}$ is the projected metric on the 3-space perpendicular to $u^a$. Also $\mu^{m}$, $p^m$, $q^{m}$ and $\pi^m_{ab}$ denotes the standard matter density, pressure, heat flux and anisotropic stress respectively. Equations (\ref{field1}) reduce to the standard Einstein field equations when $f(R) = R$.

%%%%%%%%%%%%%%%%%%%%%%%%%%%%%%%%%%%%%%%%%%%%%%%%%%%%%%%%%%%%%
\section{1+1+2 Covariant approach} \label{covariantapproach}
%%%%%%%%%%%%%%%%%%%%%%%%%%%%%%%%%%%%%%%%%%%%%%%%%%%%%%%%%%%%%
\subsection{Relativistic Cosmology} 
%%%%%%%%%%%%%%%%%%%%%%%%%%%%%%%%%%%%%%%%% 
We know that the 1+3 covariant approach, initially developed by Ehlers and Ellis \cite{bi:covariant}
has proven to be a very useful technique in many aspects of relativistic cosmology, In cosmology these methods have been applied to the formalism and evolution of density perturbations in the universe \cite{bi:pertubations} and to the physics of cosmic microwave background \cite{bi:CovCMB} amongst other things. This approach is based on a 1+3 threading decomposition of the spacetime manifold w.r.t a timelike congruence as a splitting of spacetime onto a timelike and an orthogonal three-dimensional spacelike hypersurface. All the essential information in the system is captured in a set of kinematic and dynamic 1+3 variables that have a well defined physical and geometrical significance and that satisfy a set of evolution and constraint equations derived from the Bianchi and Ricci identities, forming a closed system of equations for a chosen equation of state describing matter.

A natural extension to the 1+3 approach, optimized for problems which have spherical symmetry, is the 1+1+2 formalism developed recently by Clarkson and Barrett \cite{bi:Barrett}. In this formalism, a further splitting of the spacelike 1+3 variables is performed such that it isolates a specific spatial direction. This allows for the derivation of a set of variables that are more advantageous to treat systems with one preferred direction. For example in spherically symmetric system the equation for the 1+1+2 variables are scalar equations and are much simpler than the ones of the 1+3 formalism which are in general tensorial. 
The 1+1+2 formalism was applied to the study of linear perturbations of a Schwarzschild spacetime \cite{bi:Barrett} and to the generation of electromagnetic radiation by gravitational waves interacting with a strong magnetic field around a vibrating Schwarzschild black hole \cite{bi:Betschart}. 

In (1+3) approach first we define a timelike congruence by a timelike unit vector $u^a$. Then the spacetime is split in the form $R\otimes V$ where $R$ denotes the timeline along $u^a$ and $V$ is the 3-space perpendicular to $u^a$. Then obviously $u^au_a=-1$, and any vector $X^a$ can be projected on the 3-space by the projection tensor $h^a_b=g^a_b+u^au_b$. 

Two derivatives are defined: the vector $ u^{a} $ is used to define the \textit{covariant time derivative} (denoted by a dot) for any tensor $ T_{c..d}{}^{a..b} $ along the observers' worldlines defined by
\be
\dot{T}_{c..d}{}^{a..b} = u^{e} \nab_{e} {T}_{c..d}{}^{a..b}~, 
\ee
and the tensor $ h_{ab} $ is used to define the fully orthogonally \textit{projected covariant derivative} $D$ for any tensor $ \dot{T}_{c..d}{}^{a..b} $ , 
\be
D_{e}T_{c..d}{}^{a..b} =  h_c{}^p..h_d{}^q h_f{}^a..h_g{}^b h_e{}^r \nab_{r} {T}_{p..q}{}^{f..g}~, 
\ee
with total projection on all the free indices. 
In the (1+1+2) we further split the 3-space $V$ of the (1+3) approach, by introducing the unit vector $ e^{a} $ orthogonal to $ u^{a} $ so that
\be
e^{a} u_{a} = 0\;,\; \quad e^{a} e_{a} = 1.
\ee
Then the \textit{projection tensor}
\be 
N_{a}{}^{b} \equiv h_{a}{}^{b} - e_{a}e^{b} = g_{a}{}^{b} + u_{a}u^{b} 
- e_{a}e^{b}~,~~N^{a}{}_{a} = 2~, 
\label{projT} 
\ee 
projects vectors onto the 2-surfaces orthogonal to $e^{a}$ \textit{and} $u^a$, which we would refer as `{\it sheets}'. 

Apart from the `{\it time}' (dot) derivative, of an object (scalar, vector or tensor) which is the derivative 
along the timelike congruence $u^a$, we now introduce two new derivatives, which $e^{a}$ defines, for any object $ \psi_{a...b}{}^{c...d} $: 
\bea
\hat{\psi}_{a..b}{}^{c..d} &\equiv & e^{f}D_{f}\psi_{a..b}{}^{c..d}~, 
\\
\delta_f\psi_{a..b}{}^{c..d} &\equiv & N_{a}{}^{f}...N_{b}{}^gN_{h}{}^{c}..
N_{i}{}^{d}N_f{}^jD_j\psi_{f..g}{}^{i..j}\;.
\eea 
The hat-derivative is the derivative along the $e^a$ vector-field in the surfaces orthogonal to $u^{a}$. The $\mathds{D}$-derivative is the projected derivative onto the sheet, with the projection on every free index.
The irreducible set of kinematic and dynamical quantities:
\bea\label{irreducible}
&&\left[\Theta,~ \phi,~ \zeta_{ab},~ \A,~ \A^a,~ \alpha^a,~ a^a,~ \xi,~ \Omega,~ \Omega^a,~ \right.\\ 
&&\left. \Sigma,~ \Sigma^a,~ \Sigma_{ab},~ {\cal E},~ {\cal E}^a,~ {\cal E}_{ab},~ {\cal H},~ {\cal H}^a,~ {\cal H}_{ab} \right]
\eea
are obtained from the 1+1+2 splitting of the Weyl tensors, the covariant variation of $ e^{a} $ and the Ricci and Bianchi identities for $ e^{a} $ and $ u^{a} $. 

We see that travelling along $ u^{a} $, $ \Theta $ is the \textit{expansion scalar} (volume expansion), whilst travelling along $ e^{a} $, $\phi$ represents expansion of the sheet and $\zeta_{ab}$ is the \textit{shear of $e^{a}$} (i.e., the distortion of the sheet). $\A$ is the radial component of the acceleration of $u^{a}$, that is, of $\dot{u}^{a}$ \footnote{$\A$ can also be described as the time component of $\dot{n}^{a}$} and $\A^a$ is its component lying in the sheet orthogonal to $e^{a}$. $ \alpha^a $ is the component of $\dot{n}^{a}$ lying in the sheet and $ a^a$ is the acceleration of $ e^{a} $. $\xi$ can be interpreted as the \textit{vorticity} associated with $e^{a}$, $\Omega$ and $ \Omega^a$ are the components of the 3-vector vorticity $\omega^{a}$ along $ e^{a} $ and its sheet component respectively. The scalar, vector and tensor parts of the projected, symmetric, trace-free of the 3-tensors are $ \Sigma\,, \Sigma^a\,$ and $ \Sigma_{ab}$, respectively, for the shear,  ${\cal E}\,, {\cal E}^a\,, {\cal E}_{ab}$ for the electric Weyl tensor and ${\cal H}\,, {\cal H}^a\,, {\cal H}_{ab}$ for the magnetic Weyl tensor.

The quantities (\ref{irreducible}), together with the set of thermodynamic variables 
\be
\left[\mu, p, Q, \Pi, Q^a, \Pi_{ab}\right]\,. 
\ee
associated to a generic fluid present in the system, completely characterize the 1+1+2 spacetime. These thermodynamic quantities can be also used to represent the \textit{effective} "fluid" arising form a modification of the underlying theory of gravity. We will see that this will be actually the case in the context of fourth order gravity.

The full presentation of the 1+1+2 equations can be seen in \cite{bi:Barrett,bi:clarkson}.

%%%%%%%%%%%%%%%%%%%%%%%%%%%%%%%%%%%%%%%%%%%%%%%%%%%%%%%%%%%%%%%%%%%%%%%%
\subsection{Spherically symmetric static spacetimes in f(R) Gravity}
%%%%%%%%%%%%%%%%%%%%%%%%%%%%%%%%%%%%%%%%%%%%%%%%%%%%%%%%%%%%%%%%%%%%%%%%
LRS spacetimes posses continuous isotropy group at each point and hence a multi-transitive isometry group acting on the spacetime manifold \cite{bi:ellisLRS}. These spacetimes exhibit locally (at each point) a unique preferred spatial direction, covariantly defined, for example, by either vorticity vector field or a non-vanishing non-gravitational acceleration of the matter fluids. The 1+1+2 formalism is therefore ideally suited for covariant description of these spacetimes, yielding a complete derivation in terms of invariant scalar quantities that have physical or direct geometrical meaning \cite{bi:gary}. 
The preferred spatial direction in the LRS spacetimes constitutes a local axis of symmetry and in this case $e^{a}$ is just a vector pointing along the axis of symmetry and is thus called a 'radial' vector. Since LRS spacetimes are constructed to be locally isotropic, this allows for the vanishing of \textit{all} 1+1+2 vectors and tensors, such that there are no preferred directions in the sheet. Thus, all the non-zero 1+1+2 variables are covariantly defined scalars. The variables, 
\be
\left[\A, \Theta,\phi, \xi, \Sigma,\Omega, \E, \H, \mu, p, \Pi, Q \right]
\ee
fully describe LRS spacetimes and are what is solved for in the 1+1+2 approach. A detailed discussion of the covariant approach to LRS perfect fluid spacetimes can be found in \cite{bi:clarkson,bi:ellisLRS}.

The LRS class II spacetimes admit spherically symmetric solutions that are rotation free.  From this and the propagation and commutation equations characterizing this space time, the variables $\Omega, \xi$ and $\H$ vanish \cite{bi:gary}. The condition of staticity further implies that the dot derivatives of all the quantities vanish and consequently $\Sigma, \Theta$ and $Q$ also vanish. 

These conditions reduce the full LRS system of equations to: 
\bea \label{StSpSymEqGen}
\hat\phi &= -&\frac12\phi^2 -\frac23\mu-\frac12\Pi-\E~,
\label{equation1a}\\
\hat\E -\frac13\hat\mu + \frac12\hat\Pi &=-& \frac32\phi\bra{\E+\frac12\Pi}~,
\label{equation2a}\\
0 &= -& \A\phi + \frac13 \bra{\mu+3p} -\E +\frac12\Pi~,
\label{equation3a}\\ 
\hat p+\hat\Pi&= -&\bra{\frac32\phi+\A}\Pi-\bra{\mu+p}\A~,
\label{equation4a}\\
\hat\A &= -&\bra{A+\phi}\A + \frac12\bra{\mu +3p}~. 
\label{equation5a} 
\eea
The quantities $\mu$, $p$ and $\Pi$ are defined, in this case, as
\bea
\mu &=& \frac{1}{f,_{R}} \left(\mu^m+\frac12 ( Rf,_{R} - f) + f,_{RR}\hat X \right.\nonumber\\ \\
&&\left. + f,_{RR}X \phi+ f,_{RRR}X^{2}\right),\nonumber\\  \\
p &=& \frac{1}{f,_{R}} \left(p^m+ \frac12 ( f - Rf,_{R}) - \frac{2}{3} f,_{RR}\hat X -\right.\nonumber\\ 
&&\left.\frac{2}{3} f,_{RR}X \phi - \frac{2}{3} f,_{RRR}X^{2}- \A f,_{RR}X\right),\nonumber\\  \\
\Pi &=& \frac{1}{f,_{R}} \bra{\frac23 f,_{RRR} X^{2} + \frac23 f,_{RR} \hat X - \frac13 f,_{RR}X \phi}\,, \nonumber\\ 
\eea
where we have defined $\hat{R}=X$. We will consider the deflection angle of light propagating near an isolated spherically symmetric mass distribution so that $\mu^m=0$ and $p^m=0$. Because of the additional degrees of freedom the equations (\ref{equation1a}-\ref{equation5a}) are 
not closed and we have to add the {\it trace equation} to achieve closure: 
\be
Rf,_{R}- 2f = - 3f,_{RR}\hat X - 3f,_{RR}X \phi - 3f,_{RRR}X^{2} - 3\A f,_{RR}X \,.
\label{trace1}\\ 
\nonumber
\ee

Using the equations in (\ref{equation1a}-\ref{equation5a}) including (\ref{trace1}) and eliminating $\E$, we get the set of four coupled first order equations governing the spacetime in the fourth order gravity as
\bea
f,_{R}\bras{\hat\phi + \phi\bra{\frac12\phi-\A} }&=&\frac{1}{3}R f,_{R} - \frac23 f  \nonumber\\ 
&& +f,_{RR}X\bra{\phi + 2\A }\;, \label{eq1}\nonumber \\  \\
f,_{R}\bras{\hat\A +\A( \A + \phi)} &=& \frac16 f - 
\frac13 R f,_{R} - f,_{RR}X \A \;,\label{eq2}\nonumber \\ \\
\hat R &=& X \;, \nonumber\\ \\
f,_{RR} \hat X &=& -\frac13 Rf,_{R}+ \frac23 f 
- f,_{RRR} X^{2} \nonumber\\ 
&& - X( \phi + \A )f,_{RR} \;.\label{eq4} \nonumber\\
\eea
From the system of equations (\ref{eq1}-\ref{eq4}) we can  deduce some important results for spherically symmetric static solutions in a general $f(R)$ gravity in a completely co-ordinate independent manner (see \cite{bi:Ritu}). 

%%%%%%%%%%%%%%%%%%%%%%%%%%%%%%%%%%%%%%%%%%%%%%%%%%%%
\section{Lensing Geometry}\label{lensgeometry}
%%%%%%%%%%%%%%%%%%%%%%%%%%%%%%%%%%%%%%%%%%%%%%%%%%%%
\subsection{Null geodesics} 
%%%%%%%%%%%%%%%%%%%%%%%%%%%%%%%%%%%
We now apply the 1+1+2 approach to null geodesics characterized by a the family of null curves (or \textit{light rays}), $x^{a}(\nu)$, where $\nu$ is an affine parameter along the geodesics. The components of the tangent vector of $x^{a}(\nu)$ are
\be
k^{a} = \frac{dx^{a}}{d\nu}(\nu)
\ee
and it obeys:
\be
k^{a}k_{a} = 0~, \label{nullvector}
\ee
From this, it follows that
\be\label{geo-k}
k^{b}\na_{b}k^{a} =\frac{\delta k^{a}}{\delta \nu} = 0~, 
\ee
where we have defined $\frac{\delta }{\delta \nu} = k^{b}\na_{b}$ as the derivative along the ray. Since $ k^{a}$ is a gradient, we have that $\na_{b}k^{a} = \na_{a}k^{b}$ and this implies that \rf{geo-k} can be written as
\be\label{NullCond1} 
k_{b}\na_{a}k^{b} = 0 \,.
\ee
The light propagation vector $k^{a}$ is received by the observer from the direction determined by the the unit spatial vector $n^{a}$:
\be
n^{a}n_{a} = 1~, ~~ n^{a}u_{a} = 0
\ee
The null vector $k^a$ can be split into in the usual way \cite{bi:Ehlers}:
\be
k^a=E(u^a+n^a)\,,
\ee
A $1+1+2$  split of $n^a$ can then be performed \cite{bi:bonita}, giving
\be 
k^{a} = E(u^{a} + \kappa e^{a} + \kappa^{a})\,,\label{splitNull} 
\ee
where $E \equiv -u_{a}k^{a}$ and since $u_{a}k^{a}$ is the \textit{circular frequency} of the electromagnetic wave, $E$ can be interpreted as the energy associated with the ray. $\kappa$ is the \textit{magnitude of the radial component} and $\kappa^a$ is the \textit{component lying in the 2-dimensional sheet}. Using this form of the null vector, we can determine the lensing geometry of a photon experiencing a deflection about the centre of symmetry.

In the presence of a strong gravitational field such as a black hole, with $k^{a}$ lying tangent to the null geodesic, the general scalar deflection angle takes the form \cite{bi:bonita}: 
\be 
\alpha = \int_{\nu_{1}}^{\nu_{2}}{\frac{1}{r}\left|E\right| \sqrt{1 - \kappa^2}}d\nu - \alpha_0~. \label{DeflectionAngle} 
\ee 
This relation is geometrical and completely general. As a consequence, if we know the quantities $E(\nu)$, $\kappa(\nu)$ and $r(\nu)$, for a given spherically symmetric spacetime, it is possible to find an explicit form of the deflection angle. 

%%%%%%%%%%%%%%%%%%%%%%%%%%%%%%%%%%%%%%%%%%%%%%%%%%%%%%%%%%%%%%%%%%%%%%%%
\subsection{The propagation equations for the lensing variables.}
%%%%%%%%%%%%%%%%%%%%%%%%%%%%%%%%%%%%%%%%%%%%%%%%%%%%%%%%%%%%%%%%%%%%%%%%
Let us now look at the general propagation equations for the lensing variables $E$ and $\kappa$ in the direction of the null ray $k^a$. 
%Each ray is parameterized by the affine parameter, $\nu$, so that the geodesic condition can be written as :
%\be 
%k^{b} \nab_{b} k^{a} = \frac{\delta k^{a}}{\delta \nu} = (k^{a})' = 0~,\label{NullCond1} 
%\ee
%where the prime derivative ($'$) denotes change along the ray (i.e. with respect to the affine parameter $\nu$). 
The geodesic condition (\ref{NullCond1}) can be used to derive propagation equations for $E$ and $\kappa$. Substituting for the null vector $k^a$ (\ref{splitNull}) into (\ref{NullCond1}) and projecting the expression along the timelike direction ($u_{a}$) and along the radial direction ($e^{a}$), we obtain the general propagation equations for $E$ and $\kappa$:
\be
\frac{\delta E}{\delta \nu} =E' = - E^{2}\kappa\A- \frac32\Sigma \kappa^{2} E^{2} -E^{2} \bra{\frac13\Theta - \frac12\Sigma}~, \label{Eprime1}
\ee
\be
\frac{\delta \kappa}{\delta \nu} =\kappa{}' = E \bra{1-\kappa^{2}} \bra{\frac12 \phi - \A -\frac32\Sigma}~, \label{Kprime1}
\ee
respectively. Here, spherical symmetry has been considered and the properties
\bea
k^{b}u_{b} &=& -E,~~~ k^{b}e_{b}=E \kappa, ~~~ N^{a}{}_{b} k^{b}= E \kappa^{a},\nonumber\\
\veps^{a}{}_{b}k^{b} &=& E \veps^{a}{}_{b}\kappa^{b},~~~ u_{a}\kappa^{a} = 0,~~~ e_{a}\kappa^{a} = 0
\eea
have been used. We have also utilized the expressions
\bea
u_{a}' &=& E\A e_{a} + E\kappa \LB \frac13\Theta + \Sigma \RB
e_{a}  \\ \nonumber
&& + E\LB \frac13\Theta - \frac12\Sigma \RB\kappa_a + E\Omega \lc_{ab}\kappa^b ~,\label{uPrime} \\
e_{a}' &=& E\A u_{a} + E\kappa\LB \Sigma + \frac13\Theta \RB u_{a} +
\frac12E\phi\kappa_a + E\xi\lc_{ab}\kappa^b~, \nonumber \label{ePrime}
\eea
which are obtained from (31) and (33). In LRS spacetimes, the general propagation equations (\ref{Eprime1}) and (\ref{Kprime1}), in the direction of the ray, reduces to 
\bea 
E{}' &=& -E^{2}\A\kappa~, \label{Eprime}\\
\kappa{}' &=& E(1-\kappa^{2})(\frac12\phi-\A)~.\label{Kprime} 
\eea 
It is worth noting at this point that the form of the above equations is independent of the theory of gravity. As consequence differences in the lensing phenomenon between GR and other theories of gravity will only depend on the features of the metric of the gravitational source.

When considering spherical symmetry, it is convenient to express the hat derivative in terms of the proper radial coordinate $r$. The most natural way to do this is to make the Gaussian curvature `$K$' of the spherical sheets to be proportional to the inverse square of the radius $ K = \frac{1}{r^{2}} $. In that case, this co-ordinate `$r$' becomes the {\it area radius} of the sheets. This gives a geometrical definition to the `{\it hat}' derivative. From the propagation equation $ K $, $\hat{K} = -\phi K$ \cite{bi:gary}, the most natural way to define the hat derivative of any scalar $\Psi$ would be
\be
\hat{\Psi}=\frac{1}{2}r\phi\frac{d \Psi}{d r}\,,
\ee
for a static case. Using this we can write the propagation equations for $ E $ (\ref{Eprime}) and $ \kappa $ (\ref{Kprime}) in terms of the hat derivative as
\bea
E{}' &=& k^{a}\na_{a}E = E \kappa \hat{E} %= - E^{2} \A \kappa ~, 
\,,\\
\kappa{}' &=& k^{a}\na_{a}\kappa = E \kappa \hat{\kappa}%= E \bra{1-\kappa^{2}}\bra{\frac12\phi-\A}
\,.
\eea
Expressing the hat-derivative in terms of the radial parameter $r$ and rearranging, we obtain:
\bea
\frac{1}{E}\frac{\partial E}{\partial r} &=& - \frac{2\A}{r\phi} \label{diffE}\\
\frac{\kappa}{\bra{1-\kappa^{2}}}\frac{\partial \kappa}{\partial r} &=& \frac{1}{r}-\frac{2\A}{r\phi} \label{diffkappa}
\eea
In order to find solutions to these differential equations, one has to first  obtain solutions to the system  (\ref{eq1}-\ref{eq4}) to obtain general forms for $\A\,, \phi\,, R$ and $X$  and then substitute them into \rf{diffE} and \rf{diffkappa}. Once the solutions of \rf{diffE} and \rf{diffkappa} are found, one can substitute them into equation \rf{DeflectionAngle} to calculate the deflection angle in the presence of a strong gravitational field.

%%%%%%%%%%%%%%%%%%%%%%%%%%%%%%%%%%%%%%%%%%%%%%%%%%%%%%%%%%%%%%%%%%%%%%%%
\section{A simple example: $R^n$-gravity}\label{lensapplication}
%%%%%%%%%%%%%%%%%%%%%%%%%%%%%%%%%%%%%%%%%%%%%%%%%%%%%%%%%%%%%%%%%%%%%%%%
\subsection{Solution for the lensing variables}
%%%%%%%%%%%%%%%%%%%%%%%%%%%%%%%%%%%%%%%%%%%%%%%%%%%%
Let us now consider the simple case in which $f(R)=\chi R^n$. We will use a solution for the quantities $ \A, \phi, R $ and $ X $ found in \cite{bi:Ritu}. They read:
\begin{widetext}
\bea
\A &=& \frac{-C \bra{5 - 4 n} r^{ \frac{4 n^{2} - 11 n + 9}{n-2}}+(4 n^{2} - 6 n +2)r^{-1}}{2\bra{2-n}} \bra{\frac{\bra{1 + 2 n - 2 n^{2}} \bra{7 - 10 n + 4 n^{2}} \bra{1 + C r^{- \frac{7 - 10 n + 4 n^{2}}{2-n}}}}{(2 - n)^{2}}}^{-\frac12},\label{Adefn}\\
\phi &=& \frac{2} {r} \bra{ \frac{\bra{1 + 2 n - 2 n^{2}}\bra{7 - 10 n + 4 n^{2}}}
{\bra{2 - n}^{2} \bra{1 + C r^{- \frac{7 - 10 n + 4 n^{2}}{2-n}}}}}^{-\frac12}, \label{phidefn} \\
R &=& \frac{6n(n-1)}{\bra{2n \bra{n-1}-1}r^{2}},\label{Rdefn} \\
X &=& - \frac{12n \bra{n-1}}{\bra{2n\bra{n-1}-1}r^{3}}
\bra{\frac{\bra{1 + 2 n - 2 n^{2}}\bra{7 - 10 n + 4 n^{2}}}{\bra{2 - n}^{2} 
\bra{1 + C r^{- \frac{7 - 10 n + 4 n^{2}}{2-n}}}}}^{-\frac12}\label{Xdefn}.
\eea
\end{widetext}
It is important to remember that this metric can be used only for $n<{(1+\sqrt{3})/2}\approx 1.23$. Beyond this value of $n$ the signatures changes and the solution above must should be considered unphysical in this context. 
It is worth noting here the generalization of {\em Birkhoff's Theorem} obtained in \cite{bi:Ritu} for higher order gravity states that for all functions $f(R)$ which are of class $C^3$ at $R=0$ and $f(0)=0$ while $f,_{R}(0)\neq 0$, the Schwarzschild solution is the only static spherically symmetric vacuum  solution with vanishing Ricci scalar. This effectively implies that the form of the deflection angle for this case would be the form of the angle obtained in \cite{bi:Weinberg} for a Schwarzschild spacetime and the standard lensing results will be obtained. 
 
Now substituting for $\A$ and $\phi$ from (\ref{Adefn}) and (\ref{phidefn}) respectively in (\ref{diffE}) and solving for $E$ yields
\be
E = \frac{r^{6 + \frac{13}{n-2} + 2 n} \lambda_{1}}{\sqrt{r^{\frac{10 n}{n-2}}+ C r^{\frac{7 + 4 n^{2}}{n-2}}}}\,,
\label{Edefn}
\ee
with $ \lambda_{1} $ being an integration constant with respect to affine parameter $ \nu $. It can be determined by taking the limit of (\ref{Edefn}) as $r$ tends to some $r_{*}$. We find
\be
\lambda_{1} =E_{*}r_{*}^{-\frac{2n^{2}+2n+1}{n-2}}\sqrt{r_{*}^{\frac{10 n}{n-2}}+ C r_{*}^{\frac{7 + 4 n^{2}}{n-2}}} \label{constant1}
\ee
We are now able to solve for $\kappa$ using the differential equation for $E$. In fact using \rf{diffE} on can write:
\be
\frac{\kappa}{\bra{1-\kappa^{2}}}\frac{\partial \kappa}{\partial r}  
= \frac{1}{r} + \frac{1}{E}\frac{\partial E}{\partial r} 
\ee
This equation can be integrated giving 
\be
\kappa^{2} = \bra{1-\frac{1}{r^{2}E^{2}\lambda_{2}}}\,, 
\label{kappadefn}
\ee
where $\lambda_{2}$ is a constant of integration with respect to $\nu$. To determine $\lambda_{2}$ we consider that at 
the point of closest approach $r = r_{0}$ (i.e.  the closest point that the ray would reach in the vicinity of the lensing object), 
$ \kappa = 0$. As consequence from (\ref{kappadefn}) we have
\be
\lambda_{2} = \frac{1}{ E_{*}^{2}\bras{r_{*}^{\frac{-2(2n-1)(n-1)}{n-2}} + C r_{*}^{\frac{5-4n}{n-2}}}} \bras{\frac{r_{0}^{\frac{-4n^{2}+10 n-7}{n-2}}+C}{r_{0}^{\frac{3(2n-3)}{n-2}}}}\;. 
\label{constant2}
\ee

%%%%%%%%%%%%%%%%%%%%%%%%%%%%%%%%%%%%%%%%%%
\subsection{The Deflection Angle}
%%%%%%%%%%%%%%%%%%%%%%%%%%%%%%%%%%%%%%%%%%
We now find the form of the deflection angle (\ref{DeflectionAngle}) by substituing the solutions for $E$ (\ref{Edefn}) and 
$\kappa$ (\ref{kappadefn}) giving
\bea
\alpha = \int_{\nu_{1}}^{\nu_{2}}{\frac{E_{*}}{r^{2}}} \alpha_1 J d\nu - \alpha_{0}\,, 
\label{deflecangle}
\eea
where $J$ is the \textit{impact parameter}, given in this case by 
\be
J =\sqrt{1-\kappa^2}=\bras{\frac{r_{0}^{\frac{3(2n-3)}{n-2}}}{r_{0}^{\frac{-4n^{2}+10 n-7}{n-2}}+C}}^{\frac12}
\ee
and $\alpha_1 $ is 
\ber
\alpha_1 = r_{*}^{-\frac{2n^{2}+2n+1}{n-2}}\sqrt{r_{*}^{\frac{10 n}{n-2}}+ C r_{*}^{\frac{7 + 4 n^{2}}{n-2}}}
\ear
A transformation relation between the affine parameter $d\nu$ and radial distance $dr$ is now needed. We start with the propagation equations for $\phi$ (\ref{eq1}) applied to $R^{n}$ gravity:
\bea
\phi{}' &=& E\kappa \bra{\A-\frac{1}{2}\phi}\phi \nn\\&&+ E\kappa \bra{\frac{n-2}{3n}R + (n-1)R^{-1}X \bra{\phi +2\A}}\,.\label{Phirnprime}
\eea
Substituting in the values of the scalar functions $ \A $ (\ref{Adefn}), $ \phi $(\ref{phidefn}), $ R $ (\ref{Rdefn}), $ X $ (\ref{Xdefn}), $ E $ (\ref{Edefn}) and $ \kappa $ (\ref{kappadefn}) results in:
\begin{widetext}
\be
\phi{}'= \frac{(n-2) r^{\frac{2n^{2}-10n+5}{n-2}}\bra{2 (n-2) r^{\frac{10n}{n-2}} - C (4n^{2}-12n+11) r^{\frac{7 + 4 n^{2}}{n-2}}} \sqrt{1 - \frac{r^{ \frac{-2(n+1)(2n-3)}{n-2}} \bra{r^{2} + C r^{\frac{(2n-1)(2n-3)}{n-2}}}} {\lambda_{2} \lambda_{1}^{2}}}} {\bra{2n^{2}-2n-1} \bra{4n^{2}-10n+7} \sqrt{r^{\frac{10n}{n-2}} + C r^{\frac{7 + 4 n^{2}}{n-2}}} }r{}'\,,
\label{phiprimeclif1}
\ee
Differentiating separately the solution for $\phi$ (\ref{phidefn}) with respect to $\nu$ gives
\be
\phi{}'= \frac{-\bra{2 (n-2) r^{\frac{4+8n}{n-2}}- C\bra{4n^{2}-12n+11}r^{\frac{ 4 n^{2}-2n+11}{n-2}}}}{(n-2)\sqrt{\frac{\bra{2n^{2}-2n-1} \bra{4n^{2}-10n+7}} {(n-2)^{2} \bra{1 + C r^{\frac{4n^{2}-10n+7}{n-2}}}}}\bra{\sqrt{r^{\frac{10n}{n-2}} + C r^{\frac{7 + 4 n^{2}}{n-2}}}}} r{}' \,.
\label{phiprimeclif2}
\ee
\end{widetext}
The transformation relation is obtained by equating equations (\ref{phiprimeclif1}) and (\ref{phiprimeclif2}) giving \footnote{Note that same transformation relation is obtained when the propagation equation $ \A{}' $ and derivative of $ \A $ with respect to $ \nu $ is used in place of (\ref{phiprimeclif1}) and (\ref{phiprimeclif2}) respectively}
\be
dr = E_{*} \alpha_1 L \bras{r^{\frac{(4n^{2}-6n+2)}{n-2}}-J^{2}\bra{r^{-2}+C r^{\frac{(4n^{2}-12n+11)}{n-2}}}}^{\frac12} 
\ee
where $ L $ is
\ber
\sqrt{\frac{(n-2)^2}{(1+2n-2n^{2})(7-10n+4n^{2})}}\,.
\ear
Using this transformation relation in (\ref{deflecangle}) gives the deflection angle in the form
\bea
&&\nn \alpha =  2\int_{r_{0}}^{r_{*}} L^{-1} \frac{J}{r^{2}}\Big[r^{\frac{(4n^{2}-6n+2)}{n-2}}\\
&&- J^{2}(r^{-2}+ C r^{\frac{(4n^{2}-12n+11)}{n-2}})\Big]^{-\frac12} dr - \pi\,.
\label{deflection}
\eea
The standard form of the deflection angle in GR as given in \cite{bi:Weinberg} is recovered here when $n=1$ in (\ref{deflection}).

%%%%%%%%%%%%%%%%%%%%%%%%%%%%%%%%%%%%%%%%%%
\subsection{Observables}
%%%%%%%%%%%%%%%%%%%%%%%%%%%%%%%%%%%%%%%%%%
We now analyse the behaviour of the deflection angle $\alpha$ by computing the deflection angle against $n$  for different  distances  $ r_{*} $ from the sources and for different values of the distance of closest approach $ r_{0} $. We then plot the ratio $\alpha$ to $\alpha_{_{GR}}$, the GR case corresponding to $n=1$, against $n$ for the two cases as shown in Fig(\ref{Fig1}) and Fig(\ref{Fig2}) respectively. As a fiducial system, the distances $ r_{0} $ are in units of the Schwarzschild radius. 

The divergence of the curves in both plots is indicative of the deviation from the standard GR bending angles values. In Fig(\ref{Fig1}), the deflection angle is independent of the distance from the source $r_{*}$. This shows that the banding angle does not varies with $r_{*}$  even if the metric is not asymptotically flat.  In Fig(\ref{Fig2}) instead one can see that for a fixed $n$, the deflection angle varies for different values of $r_{0}$. In particular for $ n<1$ there is more bending as the values of $ r_{0} $ increases whereas for $ n>1$ more bending occurs as $r_{0}$ decreases. 

Fig(\ref{Fig2})  also tells us that, at fixed $r_0$  the  bending angle first decreases with $n$ for $n<1$ and then , for $n>1$, starts increasing. However it is important to remember that because of the limits on the validity of  the  \rf{Adefn}-\rf{Xdefn} our conclusions are only valid for  $n< 1.23$.

%%%%%%%%%%%%%%%%%%%%%%%%%%%%%%%%%%%%%%%%%%
\section{Conclusion}
%%%%%%%%%%%%%%%%%%%%%%%%%%%%%%%%%%%%%%%%%%
In this paper we have used the $1+1+2$ covariant approach to derive a general framework for studying strong lensing in $f(R)$ gravity. We found that the bending angle is dependent of the details of the theory of gravity, but this dependence is limited to the feature of the metric of the gravitational source. Using the simple example of $R^{n}$ gravity and one of its exact spherically symmetric solutions, we have shown that the banding angle does not depend on the position of the observer, in spite of the lack of the asymptotic flatness of the metric and that the bending angle can be increased of decreased by changing the parameters of the theory of gravity. This means that using experimental data on the strong lensing one can obtain constraints on the function $f$ and consequently obtain information on the nature of the gravitational interaction in the strong field regime. 

\section*{Acknowledgments}
SC was funded by Generalitat de Catalunya through the Beatriu de Pinos contract 2007BP-B1 00136. We thank the National Research Foundation (South Africa) for financial support. The University of Cape Town provided support by grant for RG and the National Astrophysics and Space Science Program supported AMN. 

\newpage
\begin{widetext}
\begin{figure}[h]
\label{PkRn1}
\includegraphics[scale=0.5]{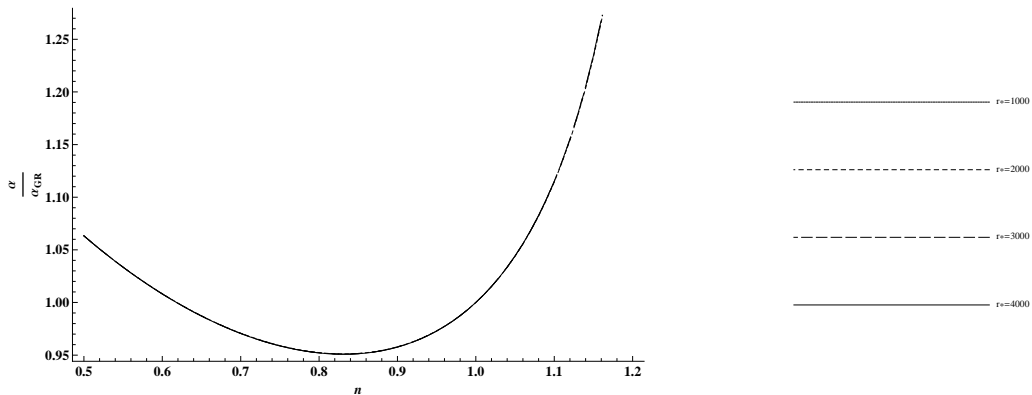}
\caption{Plot of the bending angle $\alpha$ compared to the bending angle in general relativity against $ n $ corresponding to different values of $r_{*}$, distance from the source. The result shows that the bending angle value is independent of $r_{*}$ as the plot remains the same when $r_{*}$ is varied. \label{Fig1} }
\end{figure}
\begin{figure}[h]
\label{PkRn2}
\includegraphics[scale=0.5]
{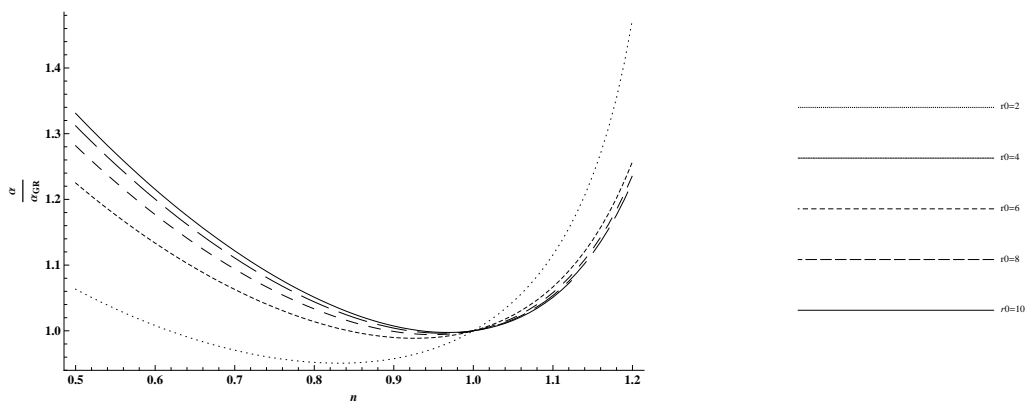}
\caption{Plot of the bending angle $\alpha$ compared to the bending angle in general relativity against $n$ corresponding to different values or $r_{0}$, distance of closest approach. \label{Fig2}}
\end{figure}
\end{widetext}

%%%%%%%%%%%%%%%%%%%%%%%%%%%%%%%%%%%%%%%%%%%%%%%%%%%%%%%

\end{document}